\documentstyle{article}
\input{psfig}

\def\beq{\begin{equation}}
\def\eeq{\end{equation}}
\def\beeq{\begin{eqnarray}}
\def\eeeq{\end{eqnarray}}
\def\to{\rightarrow}

\def\as{\alpha_S}
\def\arrowlimit#1{\mathrel{\mathop{\longrightarrow}\limits_{#1}}}
\def \pt   {\mbox{$p_{\scriptscriptstyle T}$}}
\def\cutoff{\epsilon}
\def\ltap{\raisebox{-.5ex}{\rlap{$\,\sim\,$}} \raisebox{.5ex}{$\,<\,$}}
\def\gtap{\raisebox{-.5ex}{\rlap{$\,\sim\,$}} \raisebox{.5ex}{$\,>\,$}} 

\def\np#1#2#3{Nucl.\ Phys.\ B#1 (19#3) #2}
\def\pl#1#2#3{Phys.\ Lett.\ #1B (19#3) #2}
\def\pr#1#2#3{Phys.\ Rev.\ D #1 (19#3) #2}

\def\prl#1#2#3{Phys.\ Rev.\ Lett.\ #1 (19#3) #2}

\begin{document}

\begin{titlepage}
\renewcommand{\thefootnote}{\fnsymbol{footnote}}
\begin{flushright}
     hep-ph/9610413 \\ DFF 259-10-96 \\   October 1996
     \end{flushright}
\par \vspace{10mm}
\begin{center}
{\Large \bf
Higher-order QCD corrections in hadron collisions: soft-gluon
resummation and exponentiation
}
\end{center}
\par \vspace{2mm}
\begin{center}
{\bf S. Catani}\\

\vspace{5mm}

{I.N.F.N., Sezione di Firenze}\\
{and Dipartimento
di Fisica, Universit\`a di Firenze}\\
{Largo E. Fermi 2, I-50125 Florence, Italy}
\end{center}

\par \vspace{2mm}
\begin{center} {\large \bf Abstract} \end{center}
\begin{quote}

I briefly review some recent results on soft-gluon resummation and present
applications to heavy-quark and jet production in hadron collisions.

\end{quote}
\vspace*{1cm}
\begin{flushleft}
     DFF 259-10-96 \\   October 1996
\end{flushleft}

\vspace*{2cm}

\noindent To appear in Proceedings of
{\it QCD Euroconference 96}, Montpellier, France, July 1996.

\end{titlepage}



\section{INTRODUCTION}
\label{intro}
\vspace*{-1mm}
Detailed studies on strong-interaction physics and the evaluation of the
background for new-physics signals at high-energy colliders require 
accurate calculations in perturbative QCD.
In this talk I shall describe some recent progress \cite{CMNT,topres}
in the study of a particular class of higher-order QCD contributions
to hadronic collisions. This class concerns multiple emission of soft 
gluons.

Some basic features of soft-gluon resummation are recalled in 
Sects.~\ref{soft},\ref{res}. These Sections 
provide
a brief 
introduction to the 
theoretical analysis of Ref.~\cite{CMNT}, which is 
summarised in Sect.~\ref{mpsec}. 
Phenomenological implications are
reviewed in Sect.~\ref{phen} and some  
comments are left 
to Sect.~\ref{summa}.
\vspace*{-3mm}
\section{SOFT-GLUON EFFECTS}
\label{soft}
\vspace*{-1mm}
The finite energy resolution of any particle detector
implies that physical cross
sections are always inclusive over arbitrarily-soft particles produced in the 
final state. This inclusiveness is essential in QCD calculations. Higher-order
perturbative contributions due to {\em virtual} gluons are 
infrared 
divergent: the divergences are exactly cancelled by radiation
of undetected
{\em real} gluons.

More precisely, if $1-z$ denotes the fraction of the centre-of-mass energy 
$\sqrt S$ carried by unobserved final-state particles, virtual $(v)$
and real $(r)$ soft gluons affect the cross section by the following emission
probabilities
\vspace*{-1mm}
\beeq 
\frac{dw_v(z)}{dz} \!&\!=\!&\!\! - 2a \,\delta(1-z) \int_0^{1-\cutoff} 
\frac{dz'}{1-z'}
\ln \frac{1}{1-z'} \;,\nonumber \\
\label{dw1}
\frac{dw_r(z)}{dz} \!&\!=\!&\!\! 2a \,\frac{1}{1-z} \ln \frac{1}{1-z} \, 
\Theta(1-z-\cutoff) \;,
\eeeq
where $a=C\as/\pi$ and the coefficient $C$ depends on the process.
These double-logarithmic (DL) expressions arise from the combination
of the customary bremsstrahlung spectrum $d\omega/\omega$ with
the spectrum $d\theta^2/\theta^2$ for collinear radiation.
Here I have introduced an unphysical cutoff $\cutoff$ on the minimum
energy fraction of both gluons. Adding the real and virtual terms, the physical
limit $\cutoff \to 0$ can be safely performed, leading to a finite differential
probability:
\beq
\label{dw} 
\frac{dw(z)}{dz} 
= 2a \left[ \frac{1}{1-z} \ln \frac{1}{1-z} \right]_+ \;,
\eeq
where, as usual, the notation $[ g(z) ]_+$ stands for
$\int_x^1 dz \;f(z) \,[ g(z) ]_+ \equiv  \int_x^1 dz [f(z)-f(1)] g(z) \;,$
which defines a well-behaved distribution acting on any smooth function $f(z)$
(such as cross sections or parton densities) at $z=1$.

Note that the virtual term contributes only at $z=1$ while the real one, besides
regularizing the virtual probability at $z=1$, is spread out up to the 
kinematical boundary as given by the energy fraction $x$ of the tagged final 
state. Thus the total soft-gluon contribution to the cross section is 
proportional to the following quantity
\beq
\label{dlog} 
\int_x^1 dz \frac{dw(z)}{dz} = 
-  \,a \,\ln^2 (1-x) \;\;.
\eeq
This perturbative
correction 
is the finite heritage of 
the cancellation of the infrared singularity. 

If we force the tagged final state
to carry most of the total energy (i.e. we consider the quasi-elastic limit
$x \to 1$), we strongly suppress the radiative tail of the real emission.
Thus the virtual term is unbalanced and the 
double-logarithmic 
factor in Eq.~(\ref{dlog}) can become large, $\as \ln^2(1-x) \gtap
 1$, even if the coupling constant is in the perturbative region 
$\as \ll 1$. In this kinematic regime, multiple soft-gluon radiation
produces higher-order contributions of the type
\beq
\label{logs}
C_{nm} \;\as^n \ln^m (1-x) \;, \;\;m \leq 2n \,,
\eeq
that have to be resummed in order to obtain reliable theoretical predictions.

In the case of hadron collisions, this large-$x$ region is encountered in the
production of systems of high mass $M^2$ near threshold. Outstanding examples
of these systems are the hadronic final state in deep-inelastic 
lepton-hadron scattering (DIS) (here $x=x_B$ is the Bjorken variable), lepton
pairs with large invariant mass $Q^2$ produced via the Drell-Yan (DY) mechanism
$(x\equiv \tau =Q^2/S)$ \cite{Sterman}, heavy quark-antiquark pairs 
$(x\equiv \rho = 4m_Q^2/S)$ \cite{sigres}
and pairs of jets at large transverse momentum $(x \equiv \tau = M^2_{jj}/S)$
\cite{CMNT}.

Similar soft-gluon effects 
occur
both in hadron and lepton collisions when
one measures 
detailed properties
($Q_\perp$-distributions \cite{QT},
event shapes \cite{shape}, multijet rates
\cite{CDOTW})
of the hadronic final state. 

Although the logarithmically-enhanced contributions in 
Eqs.~(\ref{dlog},\ref{logs}) are expected to be very relevant 
when
$x \to 1$, the actual size 
of the soft-gluon corrections in cross section calculations 
depends on the coefficients $C_{nm}$ and on the $x$-shape of the parton 
densities. Thus soft-gluon effects 
can be substantial also before reaching this kinematic region. This is the
main motivation 
in Refs.~\cite{sigreslaenen,BC}
for evaluating soft-gluon effects
for top quark production at Tevatron \cite{cdf,d0,top96} where the inelasticity 
variable $\rho=4m_t^2/S$ is as small as $\rho=0.04$.

\section{RESUMMATION AND EXPONENTIATION}
\label{res}
\vspace*{-1mm}
In order to discuss soft-gluon resummation in hadron collisions, let me consider
the DY process, which is the most studied process 
(see \cite{Sterman} and references therein) so far. The DY cross section
$\sigma(Q^2,\tau)$ is obtained by convoluting the parton densities 
$f(x_i,Q^2)$ of the colliding hadrons with a partonic cross section 
${\hat \sigma}(Q^2,z)$ as follows
\beq
\label{sigdy}
\sigma(Q^2,\tau)=\int_0^1 dz 
\int_0^1 dx_1 dx_2 f(x_1,Q^2) f(x_2,Q^2) 
 \;\delta(x_1x_2z-\tau) \,\Delta(z,\as(Q^2)) 
\,\sigma_0(Q^2) \;.
\eeq
Here, I have omitted all parton indices and
factorized the partonic cross section in the Born-level contribution
$\sigma_0$ times the term $\Delta$ that takes into account all the
radiative corrections. 
The latter is computable in QCD perturbation theory as a 
series in $\as$.
In Eq.~(\ref{sigdy}) I have also explicitly introduced energy 
conservation: the lepton pair carries the fraction $z=Q^2/(x_1x_2S)$ of the 
centre-of-mass energy squared $x_1x_2S$ available in the partonic subprocess. 

If the lepton pair is produced close to the hadronic threshold $(\tau \to 1)$,
energy conservation forces the partonic subprocess towards $z=1$ (as well
as the parton densities towards $x_i=1$) and the radiative corrections in 
$\Delta(z,\as)$ 
are dominated at lowest order by the soft-gluon
probability in Eq.~(\ref{dw}). The resummation of higher-order soft-gluon
effects is conveniently carried out in $N$-moment space. One can introduce the
the $N$-moments $\sigma_N(Q^2)$ of the cross section in Eq.~(\ref{sigdy})
(and likewise for any other function of longitudinal-momentum fractions) by 
performing a Mellin transformation at fixed $Q^2$:
\vspace*{-1mm}
\beq
\label{sign}
\sigma_N(Q^2) \equiv \int_0^1 d\tau \;\tau^{N-1} \;\sigma(Q^2,\tau) \;.
\eeq
\vspace*{-1mm}
Working in $N$-moment space, any $z$-distribution is replaced by a function of
$N$ and, due to the weight factor $\tau^{N-1}$, the threshold region is sampled 
by the limit $N \to \infty$. However, the main reason for using $N$-moments is
not the replacement of distributions by functions, but rather the fact that
one can easily implement momentum conservation. For instance, the 
energy-conservation constraint in Eq.~(\ref{sigdy}) is exactly diagonalized and
the $N$-moments of the cross section have the following factorized expression
\beq
\label{sigdyn}
\!\sigma_N(Q^2) = f_N(Q^2) f_N(Q^2) \Delta_N(\as(Q^2)) \,\sigma_0(Q^2) \,.
\eeq

This kinematical factorization plays an essential role in the resummation
programme.  
The radiative
factor $\Delta(z,\as)$ is obtained as follows 
\vspace*{-1.5mm}
\beq
\label{dres}
\Delta(z,\as) = \delta(1-z) + \sum_{n=1}^{+\infty}
\int_0^1 dz_1 \dots dz_n  \; \frac{dw_n(z_1,\dots,z_n)}{dz_1 \dots dz_n}
\; \Theta_{{\rm PS}}(z;z_1,\dots,z_n) \;,
\eeq
where the probability $dw_n$ for producing $n$ soft gluons is integrated over 
the phase-space region (symbolically denoted by $\Theta_{{\rm PS}}$) that is 
available according to the actual definition of the cross section.
The first relevant ingredient to perform the summation in 
Eq.~(\ref{dres}) has a dynamical origin. Owing to the factorization properties
of multi-gluon QCD amplitudes in the soft limit, the $n$-gluon probability
factorizes in the product of the single-gluon contributions $dw(z_i)$ 
in Eq.~(\ref{dw}):
\beq
\label{wfac}
\frac{dw_n(z_1,\dots,z_n)}{dz_1 \dots dz_n} \simeq \frac{1}{n!} 
\prod_{i=1}^{n} \frac{dw(z_i)}{dz_i} \;.
\eeq
The second relevant information for the summation regards the kinematics. 
In general, the phase-space function $\Theta_{{\rm PS}}$ depends in a 
non-trivial way on the multi-gluon configuration so that it
cannot be handled in a simple manner. However, 
in the case of total cross sections, like in the DY process, the only relevant
constraint is longitudinal-momentum conservation. This constraint is
exactly factorizable going to $N$-space: 
\beq
\label{tfac}
\Theta_{{\rm PS}}(z;z_1,\dots,z_n) = \delta(z - z_1 \dots z_n) 
\; \arrowlimit{N-{\rm moments} } z_1^{N-1} \dots z_n^{N-1}
\eeq
Using Eqs.~(\ref{wfac}) and (\ref{tfac}) one can straightforwardly recast 
Eq.~(\ref{dres}) in exponential form:
\beq
\label{dnres}
\Delta_N(\as) = 
\exp \left[ \,\int_0^1 dz  \,z^{N-1} \frac{dw(z)}{dz} \right] \;.
\eeq
In summary, the physical basis for the exponentiation of soft-gluon 
contributions are the factorization of both the multi-gluon amplitudes 
and the phase-space. The first property follows from QCD dynamics and is 
completely general. The second property depends on the actual  
definition of the cross section and can be easily violated even in QED 
\cite{BS}. In the case of total cross sections in hadron collisions, because of 
energy conservation,
phase-space factorization can only be achieved by working in $N$-space. 
There are no physical basis for exponentiation directly in $x$-space.

Inserting Eq.~(\ref{dw}) into Eq.~(\ref{dnres}) and performing the large-$N$ 
limit, one finds the exponential of a DL
expression:
\beeq
\label{dndlz}
\!\!\Delta_N(\as) &\!\!\!\!=\!\!\!& 
\exp \left[ \,2a \int_0^1 dz \frac{z^{N-1}-1}{1-z} 
\ln\frac{1}{1-z} \right]  \\
\label{dndl}
&\!\!\!\!=\!\!\!& \exp 
\left[ - a \;( \,\ln^2 N + {\cal O}(\ln N) ) \right] \;.
\eeeq
The derivation of Eq.~(\ref{dndl}) that I have 
sketched would be complete
in massless QED (neglecting the effect of soft-fermion pairs). The only 
simplification I have introduced in the QCD case regards the factorization
formula (\ref{wfac}). Unlike photons, which have no electric charge, gluons
carry colour charge and multiple gluon emission is affected by dynamical
correlations. Nonetheless, using general properties related to gauge invariance 
and unitarity, one can prove that, in the DY process and similar total cross
sections, a generalized exponentiation theorem is valid in the following form  
\cite{Sterman}
\vspace*{-1mm}
\beq
\label{dngen}
\ln \Delta_N(\as) 
= \ln N \,g_1(\as \ln N) + \,g_2(\as \ln N) 
+ \as \; g_3(\as \ln N) 
+ \dots \;.
\eeq
Since the functions $g_i$ 
depend only on $\as \ln N$,
in this context exponentiation means that all the logarithmic corrections
of the type $\as^n \ln^m N$ with $n + 1 < m \leq 2n$, 
allowed by
Eq.~(\ref{logs}), 
are simply taken into account by the exponentiation of 
lowest-order terms and thus cancel in the exponent of $\Delta_N$. Owing to the 
general structure of Eq.~(\ref{dngen}), one can control the resummation of the
logarithmically-enhanced terms by introducing an improved perturbative 
expansion. The function $g_1$ gives the {\em leading} logarithmic (LL) 
contributions $\as^n \ln^{n+1} N$, $g_2$ contains the {\em next-to-leading}
logarithmic (NLL) terms $\as^n \ln^{n} N$, 
etc.

The functions $g_i$ essentially represent the QCD correction to the 
factorization in Eq.~(\ref{wfac}). Most of 
the correction is simply taken into account by introducing running coupling 
effects. Roughly speaking, this amounts to the replacement 
$a=C\as(Q^2)/\pi \to C\as((1-z)Q^2)/\pi$ at the integrand level in 
Eq.~(\ref{dndlz}). Thus, because of the logarithmic behaviour
$\as(\mu^2) \sim 1/\ln (\mu^2/\Lambda^2)$, integrating exactly in $z$ up
to $z=1$ one eventually hits the Landau pole at $z=1-\Lambda^2/Q^2$.
This singularity, called infrared renormalon \cite{BB,zak}, 
signals the onset of non-perturbative phenomena and has to be properly 
regularized.

\section{IMPLEMENTING RESUMMED FORMULAE IN HADRON COLLISIONS:
THE MINIMAL PRESCRIPTION}
\label{mpsec}
\vspace*{-1mm}
The theoretical analysis in Ref.~\cite{CMNT} deals with difficulties that 
arise when one tries to apply resummed formulae 
of the type discussed in Sect.~\ref{res}
to the actual evaluation of physical cross sections.

The first difficulty is due to the fact that these formulae
are provided
in $N$-space. Cross sections like that in Eq.~(\ref{sigdy})
are not experimentally
measurable in the entire kinematic range $0 \leq \tau \leq 1$ 
and thus theoretical predictions for the 
$N$-moments cannot directly be compared with data.
In any phenomenological application one has to go back to $x$-space and 
compute the radiative factor $\Delta(z,\as)$.

The regularization of the running coupling singularity noticed at the end
of Sec.~\ref{res} leads  
to the second difficulty. One has to introduce a regularization 
prescription that does not spoil the main perturbative features.

As discussed in Ref.~\cite{CMNT}, several methods used in the literature
to solve these difficulties are theoretically unjustified 
and, typically, enhance soft-gluon effects long 
before the hadronic-threshold region is actually approached.

In particular, we pointed out the differences between the above difficulties and
we warned of the danger of exploiting the formal equivalence
$N \sim 1/(1-x)$ to straightforwardly transform $N$-space resummed formulae
into $x$-space resummed formulae. For instance, following 
Refs.~\cite{sigres,sigreslaenen,BC}, one would replace the DL expression 
(\ref{dndl}) with the $x$-space version:
\vspace*{-1mm}
\beq
\label{dlz}
\Delta(z,\as) \simeq - \frac{d}{dz} \left[ \Theta(1-z-\cutoff) 
\;e^{-a\ln^2(1-z)} \; 
(1+ \dots) \right] \,,
\eeq
where the dots stand for `subleading' logarithmic terms, i.e. terms at most of
the order of $\as^n \ln^n(1-z)$. Expanding Eq.~(\ref{dlz}) as a power series in 
$\as$ and identifying $\cutoff$ with the unphysical cutoff in Eq.~(\ref{dw1}), 
one obtains the lowest-order 
probability in Eq.~(\ref{dw}). Equation (\ref{dlz}) is
formally valid \cite{shape} in terms of the logarithmic expansion in 
$\ln (1-z)$, but the identification  $\ln N \to \ln 1/(1-z)$ and the ensuing
approximation of neglecting subleading logs is correct only if 
$\as \ln 1/(1-z) \ltap 1$. Care has to be taken before extending this 
approximation to higher values of $z$.

In particular, it is quite dangerous to insert Eq.~(\ref{dlz}) into the 
factorization formula (\ref{sigdy}), where the $z$-integration range extends up
to $z=1$. In the case of hadron collisions, after having introduced the parton
densities $f(x_i,Q^2)$ in customary factorization schemes (like the DIS or
${\overline {\rm MS}}$ schemes), the coefficient $C$ in $a=C\as/\pi$ turns out
to be {\em negative}. Thus the expression 
$\exp [-a\ln^2(1-z)] = \exp [|a|\ln^2(1-z)]$
diverges faster than any power of $(1-z)$ and is not integrable at $z=1$:
in the resummed expression (\ref{dlz}) the unphysical cutoff $\cutoff$ can no
longer be removed without producing a singularity. Note that such a singularity
is not related to the hadronic threshold. The inelasticity variable $\tau$
controls the lower limit of the $z$-integral in Eq.~(\ref{sigdy}): the point
$z=1$ is inside the integration region even asymptotically far $(\tau \to 0)$
from threshold.
 
The origin of this singularity was studied in detail in Ref.~\cite{CMNT} 
(see also \cite{PN})
and traced back to the {\em same-sign} factorial divergence of 
the `subleading' terms
systematically neglected in the derivation of the $x$-space formula (\ref{dlz})
from Eq.~(\ref{dndl}). No analogous divergence is present in the original 
$N$-space formula.
Soft-gluon exponentiation in $N$-space is related to
momentum conservation. Using exponentiation directly in $x$-space, one violates
infinitely many times momentum conservation and can build up a (kinematical) 
divergence.

In Ref.~\cite{CMNT} we specified a 
prescription for the implementation of soft-gluon resummation formulae in the 
computation of cross sections in hadronic collisions. We used the generalized 
exponentiation
formula in Eq.~(\ref{dngen}) and performed numerically the inversion from
$N$-space to $x$-space without further approximations. 
As for the regularization of the 
Landau pole in the running coupling, we pointed out that the functions 
$g_i(\as \ln N)$ are well-defined analytic functions of $N$ in the complex 
plane. The Landau pole only produces a branch point at $N=Q/\Lambda$. In the 
numerical inversion we used a prescription that leaves this non-perturbative
branch cut to the right of the integration contour.
  
As proved in Ref.~\cite{CMNT}, with our prescription 
no factorial divergence is introduced
in the resummed 
expansion. Factorially growing terms 
related to infrared renormalons are likely to
be present in the full perturbative QCD series, but as shown in 
Ref.~\cite{BB}, their identification and evaluation require an analysis whose
accuracy goes beyond that used to prove the exponentiation of 
logarithmically-enhanced contributions. Since our prescription fulfils
the relevant kinematical constraints and does not introduce large corrections
that are not justified by the logarithmic expansion, we call it the
`Minimal Prescription' (MP). This prescription can be used to extend 
perturbative QCD calculations towards the threshold region by consistently 
taking into account logarithmically-enhanced soft-gluon effects.

\section{PHENOMENOLOGICAL STUDIES}
\label{phen}
\vspace*{-1mm}
In Ref.~\cite{CMNT}, the MP 
was applied
to the
DY process and to the production of heavy quarks and 
high-$\pt$ jets in hadron collisions. Since the last two processes
are becoming increasingly topical, in this Section I briefly review our results.
In both cases, in the resummation of soft-gluon corrections we included
only the LL contributions (i.e. the analogue of the function $g_1$ in 
Eq.~(\ref{dngen})). In the case of heavy-quarks, this corresponds
(roughly speaking) to the $N$-space version
of the resummation considered in Ref.~\cite{sigres}.  

\subsection{Heavy-quark production}
\label{hqsec}

The importance of the resummation effects for top quark production in
$p{\bar p}$-collisions was studied by computing the following quantities
\beq \label{deltadef} \frac{\delta_{\rm
    gg}}{\sigma^{(gg)}_{\rm NLO}}\,,\quad \frac{\delta_{\rm
    q\bar{q}}}{\sigma^{(q\bar{q})}_{\rm NLO}}\,,\quad
\frac{\delta_{\rm gg}+\delta_{\rm q\bar{q}}}{\sigma^{(gg)}_{\rm
    NLO}+\sigma^{(q\bar{q})}_{\rm NLO}}\;,
\eeq
that are plotted in Fig.~1.
Here $\delta_{ab}$ is equal
to our MP-resummed hadronic cross section in which the terms of order
$\as^2$ and $\as^3$ have been subtracted, and $\sigma_{(\rm NLO)}^{(ab)}$ is
the hadronic cross section in full next-to-leading order \cite{sigtot} (i.e.
including both soft and hard radiation up to order $\as^3$). The labels $ab$
refer to the various partonic channel that contribute to the cross sections.

At Tevatron energies one can see that the contribution of resummation
is very small, being of the order of 1\% for $m_t\simeq 175 {\rm GeV}$
(in the calculation of Ref.~\cite{BC} the resummation effect is about 10\%,
i.e. one order of magnitude higher; similar quantitative effects are found
with the method of Refs.~\cite{sigres,sigreslaenen}). Thus, as expected 
by the small value of $\rho=4m_t^2/S \sim 0.04$, we concluded that top
quark production at Tevatron is reliably estimated by the next-to-leading
order (NLO) QCD calculation. Based upon these findings, we 
updated the computation of Ref.~\cite{gual}. Full details of our
calculation are given in Ref.~\cite{topres}.
The QCD prediction is conveniently
parametrized as follows
\beq
\sigma_{t{\bar t}}(1.8 \,{\rm TeV})=e^{\frac{175-m_{\rm t}}{31.5}}
(\mbox{$4.75{+0.73\atop-0.62}$}) \,{\rm pb}\,.
\eeq

We evaluated others heavy-quark production cross sections.
In general we found 
\cite{CMNT}
that, in most experimental configurations of practical
interest, soft-gluon resummation effects are not dominant either because
they are very small or because they are well below the (estimated)
uncertainty due to higher-order (non-soft) 
corrections.
\begin{figure}
\centerline{\psfig{figure=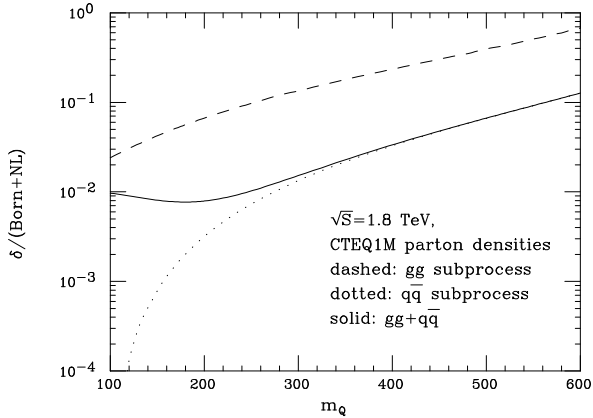,width=0.48\textwidth,clip=}}
\caption{\label{frtev}
Contribution of gluon resummation at order $\as^4$ and higher, relative to the
NLO result, for the individual subprocesses and for the 
total,
as a function of the top mass in $p\bar p$ collisions at 1.8 TeV. }
\end{figure}
\begin{figure}
\centerline{\psfig{figure=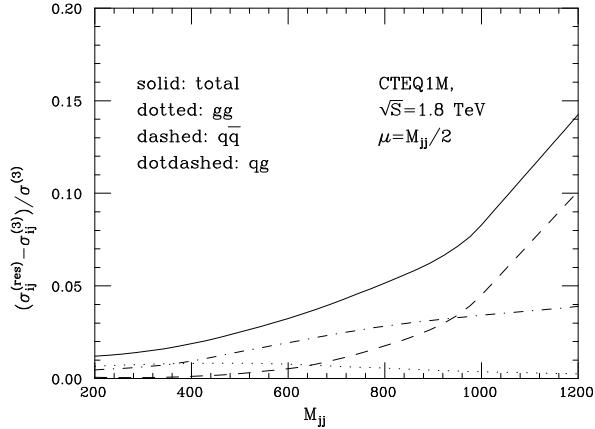,width=0.48\textwidth,clip=}}
\caption{\label{jetcteq2}
Contribution of gluon resummation at order $\as^4$ and higher, relative to the
truncated ${\cal O}(\as^3)$ result,
for the invariant-mass distribution of jet pairs at the Tevatron.}
\end{figure}

\subsection{Jet production}
\label{jetsec}

The interest in the effects of soft-gluon resummation on the behaviour of jet 
cross sections 
is prompted by the discrepancy 
between the
single-inclusive jet distribution at large \pt, as measured by CDF
\cite{cdfjet}, and the result of the NLO QCD predictions \cite{nlojet}.  
This discrepancy is the topic of 
ongoing investigations and discussions \cite{etdisc} on both experimental and 
theoretical aspects.

Owing to its kinematic similarity with DY and heavy-flavour production,
we studied \cite{CMNT} the soft-gluon corrections to the 
invariant-mass distribution of a jet pair. For this distribution 
an analogous discrepancy between data and theory
has been reported \cite{cdfmass}.
Figure~2
shows our results for the following quantities
\beq
\frac{\delta^{(4)}_{\rm gg}}{\sigma^{(3)}}\,,\quad
\frac{\delta^{(4)}_{\rm qg}}{\sigma^{(3)}}\,,\quad \frac{\delta^{(4)}_{\rm
    q\bar{q}}}{\sigma^{(3)}}\,,\quad \frac{\delta^{(4)}_{\rm
    gg}+\delta^{(4)}_{\rm qg}+\delta^{(4)}_{\rm q\bar{q}}}{\sigma^{(3)}} \;,
\eeq
where, analogously to Eq.~(\ref{deltadef}),
$\delta^{(4)}$ is equal to the MP resummed hadronic cross section
with terms of order $\as^3$ subtracted. Unlike in Eq.~(\ref{deltadef}),
$\sigma^{(3)}$ is not the full NLO cross section but rather its approximation
as obtained by truncating the resummation formula at order $\as^3$.  
One can see that the resummation effect leads to 
$10\div15\%$ increase of the cross section 
when the inelasticity variable $\tau=M_{jj}^2/S$ is as large as $\tau \sim 0.5$.

These results
should only be taken as an indication of the order of magnitude of the
correction. This is because we did not include a study of the resummation
effects on the determination of the parton densities, we only considered
LL resummation and we did not implement detailed experimental cuts.

Note also that other distributions, such as the $\pt$ of the jet, 
have a rather different
structure from the viewpoint of soft-gluon resummation because of additional
jet-broadening effects due to final-state radiation. Calculations of these
effects are in progress.

\section{SUMMARY AND OUTLOOK}
\label{summa}

The MP can be used to evaluate the effect of soft-gluon resummation on cross
sections in hadron collisions. Using the MP we find that the 
logarithmically-enhanced soft-gluon terms become important only fairly
close to the hadronic threshold $(x \sim 0.5)$. Further investigations of
resummation effects in this kinematic region are warranted. These include
the consistent determination and $Q^2$-evolution of the parton densities
by using resummed anomalous dimensions \cite{Sterman} and the evaluation
of NLL corrections \cite{SNLL}.

\vspace*{3mm}
\noindent{\bf Acknowledgements}. This research is supported in part by 
EEC Programme `Human Capital and Mobility', Network 
`Physics at High Energy Colliders',
contract CHRX-CT93-0357 (DG 12 COMA).
I would like to thank Stephan Narison for his continual effort in the 
succesful organization of this Conference.



\begin{thebibliography}{9}

\bibitem{CMNT}
  S.\ Catani, M.L.\ Mangano, P.\ Nason and L.\ Trentadue, 
  \np{478}{273}{96}.
  
\bibitem{topres}
  S.\ Catani, M.L.\ Mangano, P.\ Nason and L.\ Trentadue, \pl{378}{329}{96}.

\bibitem{Sterman}
  G.\ Sterman, \np{281}{310}{87};
  S.\ Catani and L.\ Trentadue, \np{327}{323}{89},
  \np{353}{183}{91}. 

\bibitem{sigres}
  E.\ Laenen, J.\ Smith and W.L.\ van Neerven,
  \np{369}{543}{92}. 

\bibitem{QT} 
J.\ Kodaira and L.\ Trentadue, \pl{123}{335}{82}; J.C.\ Collins and 
D.E.\ Soper, \np{197}{446}{82};
G.\ Altarelli, R.K.\ Ellis, M.\ Greco, G.\ Martinelli, \np{246}{12}{84};
J.C.\ Collins, D.E.\ Soper and G. Sterman, \np{250}{199}{85}.

\bibitem{shape}
S.\ Catani, G.\ Turnock, B.R.\ Webber and L.\ Trenta\-due, \pl{263}{491}{91},
\np{407}{3}{93}.

\bibitem{CDOTW}
S.\ Catani, Yu.L.\ Dokshitzer, M.\ Olsson, G.\ Tur\-nock and B.R.\ Webber,
\pl{269}{432}{91}.

\bibitem{sigreslaenen}  
  E.\ Laenen, J.\ Smith and W.L.\ van Neerven,
  \pl{321}{254}{94}.
  
\bibitem{BC}
  E.L.\ Berger and H.\ Contopanagos, \pl{361}{115}{95},
  \pr{54}{3085}{96}, 
  ANL-HEP-CP-96-51 (hep-ph/9606421),
  to appear in Proc. of 11th Topical
  Workshop on Proton-Antiproton Collider Physics (PBARP 96).

\bibitem{cdf}
  CDF Coll., F.\ Abe et al., \pr{50}{2966}{94},
  \prl{74}{2626}{95}.

\bibitem{d0}
  D0 Coll., S.\ Abachi et al., \prl{74}{2632}{95}.

\bibitem{top96} 
  S.\ Leone, these proceedings.   

\bibitem{BS}
N.\ Brown and W.J.\ Stirling, Phys. Lett. 252B (1990) 657;
S.\ Catani, in {\it QCD at 200 TeV}, Proc. 17th Eloisa\-tron Project Workshop,
eds. L.\ Cifarelli and Yu.L.\ Dokshitzer (Plenum Press, New York, 1992),
p.~21.

\bibitem{BB}
  M.\ Beneke and V.M.\ Braun, \np{454}{253}{95};
  Yu.L.\ Dokshitser, G.\ Marchesini and B.R. Webber, \np{469}{93}{96};
  M.\ Beneke, SLAC-PUB-7277 (hep-ph/9609215), to appear in Proc.
  28th International Conference on High-energy Physics (ICHEP 96).

\bibitem{zak}
  V.I.\ Zakharov, these proceedings.

\bibitem{PN}
P.~Nason, CERN-TH/96-204 (hep-ph/9607430), to appear in Proc. of
the 10th Rencontres de Physique de la Vall\'{e}e d'Aoste.


\bibitem{sigtot}
  P.\ Nason, S.\ Dawson and R.K.\ Ellis,
  \np{303}{607}{88};
  W.\ Beenakker, H.\ Kuijf, W.L.\ van Neerven and J.\ Smith,
  \pr{40}{54}{89}.

\bibitem{gual}
  G.\ Altarelli, M.\ Diemoz, G.\ Martinelli and P.\ Nason,
  \np{308}{724}{88};
  R.K.\ Ellis, \pl{259}{492}{91}.

\bibitem{cdfjet}
 CDF Coll., F.\ Abe et al., \prl{77}{438}{96}.

\bibitem{nlojet}
   F.\ Aversa, P.\ Chiappetta, M.\ Greco and J.P.\ Guillet,
   \np{327}{105}{89}, \prl{65}{401}{90};
   S.\ Ellis, Z.\ Kunszt and D.\ Soper,
    \pr{40}{2188}{89}, \prl{64}{2121}{90};
   W.T.\ Giele, E.W.N.\ Glover and D.A.\ Kosower, \np{403}{633}{93}. 

\bibitem{etdisc}
See references in the contributions by S.\ Grunendhal, R.\ Plunkett
and D.A.\ Soper to these proceedings.

\bibitem{cdfmass}
  E.~Buckley-Geer, for the CDF Collaboration, 
  FERMILAB-CONF-95-316-E, to appear in Proc. of Int. Europhysics
  Conference on High
  Energy Physics (HEP 95).

\bibitem{SNLL}
  N.\ Kidonakis and G.\ Sterman, ITP-SB-96-7 (hep-ph/9604234); 
  H.\ Contopanagos, E.\ Laenen and G.\ Sterman, ANL-HEP-25
  (hep-ph/9604313).

\end{thebibliography}
\end{document}